\documentclass[aps,prd,reprint,superscriptaddress,nofootinbib]{revtex4-2}

\usepackage{graphicx}% Include figure files
\usepackage{dcolumn}% Align table columns on decimal point
\usepackage{bm}% bold math
\usepackage{amsmath}
\usepackage{hyperref}
\usepackage{natbib}
\usepackage{subcaption}
\begin{document}

%\preprint{APS/123-QED}

\title{Projected Sensitivity to Slow Muonphilic Dark Matter with Accelerator Muon Beams}% Force line breaks with \\

\author{Rongfeng Zhang}
\affiliation{%
 State Key Laboratory of Nuclear Physics and Technology, School of Physics, Peking University\\
 Beijing 100871, China
}

\author{Cheng-en Liu}
\affiliation{%
 State Key Laboratory of Nuclear Physics and Technology, School of Physics, Peking University\\
 Beijing 100871, China
}

\author{Ruihu Zhu}
\affiliation{%
 Institute of Modern Physics, Chinese Academy of Sciences\\
 Lanzhou 730000, China
}
\affiliation{%
 University of Chinese Academy of Sciences\\
 Beijing 100049, China
}

\author{Yu Xu}
\affiliation{%
 Institute of Modern Physics, Chinese Academy of Sciences\\
 Lanzhou 730000, China
}
\affiliation{%
 Advanced Energy Science and Technology Guangdong Laboratory\\
 Huizhou 516000, China
}

\author{Zijian Wang}
\affiliation{%
 State Key Laboratory of Nuclear Physics and Technology, School of Physics, Peking University\\
 Beijing 100871, China
}

\author{Leyun Gao}
\affiliation{%
 State Key Laboratory of Nuclear Physics and Technology, School of Physics, Peking University\\
 Beijing 100871, China
}

\author{Xueheng Zhang}
\affiliation{%
 Institute of Modern Physics, Chinese Academy of Sciences\\
 Lanzhou 730000, China
}
\affiliation{%
 Advanced Energy Science and Technology Guangdong Laboratory\\
 Huizhou 516000, China
}
\affiliation{%
 School of Nuclear Science and Technology, University of Chinese Academy of Sciences\\
 Beijing 100049, China
}

\author{Qite Li}
\email{Contact author: liqt@pku.edu.cn}
\affiliation{%
 State Key Laboratory of Nuclear Physics and Technology, School of Physics, Peking University\\
 Beijing 100871, China
}

\author{Liangwen Chen}
\affiliation{%
 Institute of Modern Physics, Chinese Academy of Sciences\\
 Lanzhou 730000, China
}
\affiliation{%
 Advanced Energy Science and Technology Guangdong Laboratory\\
 Huizhou 516000, China
}
\affiliation{%
 School of Nuclear Science and Technology, University of Chinese Academy of Sciences\\
 Beijing 100049, China
}

\author{Chen Zhou}
\email{Contact author: czhouphy@pku.edu.cn}%
\affiliation{%
 State Key Laboratory of Nuclear Physics and Technology, School of Physics, Peking University\\
 Beijing 100871, China
}

\author{Qiang Li}
\email{Contact author: qliphy0@pku.edu.cn}%
\affiliation{%
 State Key Laboratory of Nuclear Physics and Technology, School of Physics, Peking University\\
 Beijing 100871, China
}

\author{Zhiyu Sun}
\affiliation{%
 Institute of Modern Physics, Chinese Academy of Sciences\\
 Lanzhou 730000, China
}
\affiliation{%
 Advanced Energy Science and Technology Guangdong Laboratory\\
 Huizhou 516000, China
}
\affiliation{%
 School of Nuclear Science and Technology, University of Chinese Academy of Sciences\\
 Beijing 100049, China
}

\begin{abstract}
The nature of dark matter (DM) remains one of the most enduring open questions in modern physics, and muonphilic DM has emerged as a promising scenario that complements traditional DM candidates.  Following the recently established cosmic-ray muon scattering approach, we investigate the sensitivity for probing slow muonphilic DM with accelerator muon beams. A Geant4-based simulation framework is developed, incorporating the detector geometry from the PKMu muon tomography system and a dedicated elastic $\mu$-DM scattering process. The projected sensitivity is found to be largely insensitive to both the beam energy and the transverse beam size when the beam is fully contained within the detector acceptance. For a benchmark beam intensity of $10^5/\rm{s}$, the simulated pure-muon beam surpasses the existing cosmic-ray limit of $1.61\times10^{-17}$ cm$^2$ at $m_{\rm DM}=1$ GeV within approximately 11 seconds. A realistic muon beam phase-space distribution based on simulations for the High Intensity heavy-ion Accelerator Facility (HIAF) is also implemented, yielding projected limits that improve upon the cosmic-ray results by nearly two orders of magnitude in a one-day exposure. These results demonstrate that a beam-muon scattering experiment offers a robust and promising route toward significantly improved sensitivity to slow muonphilic DM.

\end{abstract}

%\keywords{Suggested keywords}%Use showkeys class option if keyword
                              %display desired
\maketitle

%\tableofcontents

\section{Introduction}

Dark matter (DM) is one of the most compelling pieces of evidence for physics beyond the standard model (SM). Cosmological and astrophysical observations have established the existence of a nonluminous matter component, while its microscopic nature remains unknown \citep{2020, BERTONE2005279}. The absence of conclusive signals in conventional searches has motivated a broad experimental program targeting light DM candidates, subdominant strongly interacting relics, and nonstandard portals to the SM \cite{essig2023snowmass2021cosmicfrontierlandscape}. Among these possibilities, leptophilic or muon-specific interactions are particularly interesting, since they naturally evade part of the constraints from nuclear-recoil direct-detection experiments and can be explored with precision muon measurements and accelerator-based muon beams \cite{Bai2014}.

Muon beams provide a clean and powerful probe of such scenarios. Recent and proposed experiments, such as NA64$\mu$ and MUonE, demonstrate that high-intensity muon beams can be used to search for dark sectors and to perform precision scattering measurements \citep{PhysRevLett.132.211803, CARLONICALAME2015325}. The NA64$\mu$ experiment at the CERN SPS, operating with a 160 GeV muon beam, has established competitive constraints on light dark matter and $L_\mu-L_\tau$ gauge bosons, with a data sample of approximately $3.5\times10^{11}$ muons-on-target collected during the 2023–2024 runs \citep{NA64_2024}. The MUonE experiment, also at CERN's M2 beam line, aims to measure the hadronic contribution to the muon anomalous magnetic moment through high-precision muon-electron elastic scattering \cite{MUonE_2024}. Most accelerator searches focus on the production of invisible particles through missing-energy or missing-momentum signatures. 

A complementary possibility is to use muons as incident probes of an ambient slow DM population. In this case, if DM couples preferentially to muons, elastic $\mu$-DM scattering may induce a measurable deflection of the muon trajectory. This approach is especially relevant for strongly interacting relics that may be captured and thermalized inside Earth, leading to a large near-surface density enhancement for a subcomponent of DM \citep{PhysRevD.103.115031, PhysRevLett.131.011005, PhysRevD.109.075027}. A first experimental realization of this idea was recently achieved by the PKMu (Probing and Knocking with Muons) collaboration \cite{5jh7-fxf4, PKMu_PRD2024, Gao2025}, using cosmic-ray muons and a resistive plate chamber (RPC) tomography system. In that study, the scattering angle between incident and outgoing cosmic-ray tracks was introduced as the key observable. A 63-day data-taking campaign recorded 1.18 million effective scattering events; the data were used to extract the sea-level secondary cosmic-ray composition and to constrain slow muonphilic DM. For a 1 GeV DM particle, the 95\% confidence-level upper limit on the elastic $\mu$-DM scattering cross section reached $1.61 \times 10^{-17}$ $\rm{cm}^{2}$, assuming an Earth-bound density enhancement scenario \cite{5jh7-fxf4}. This result established the feasibility of searching for muonphilic DM through angular scattering measurements. Beyond the DM search, the PKMu framework has also been extended to probe charged lepton flavor violation (CLFV) \cite{Gao2024_CLFV}, to explore dark photon and dark mediator signatures in muon-electron scattering \cite{arXiv:2607.18669}, and to investigate quantum entanglement and Bell inequality violation in GeV-scale muon-electron scattering \cite{Ruzi2025_QE}. These developments highlight the versatility of the muon scattering approach for probing new physics beyond the standard model.

\begin{figure}
    \centering
    \includegraphics[width=1.0\columnwidth]{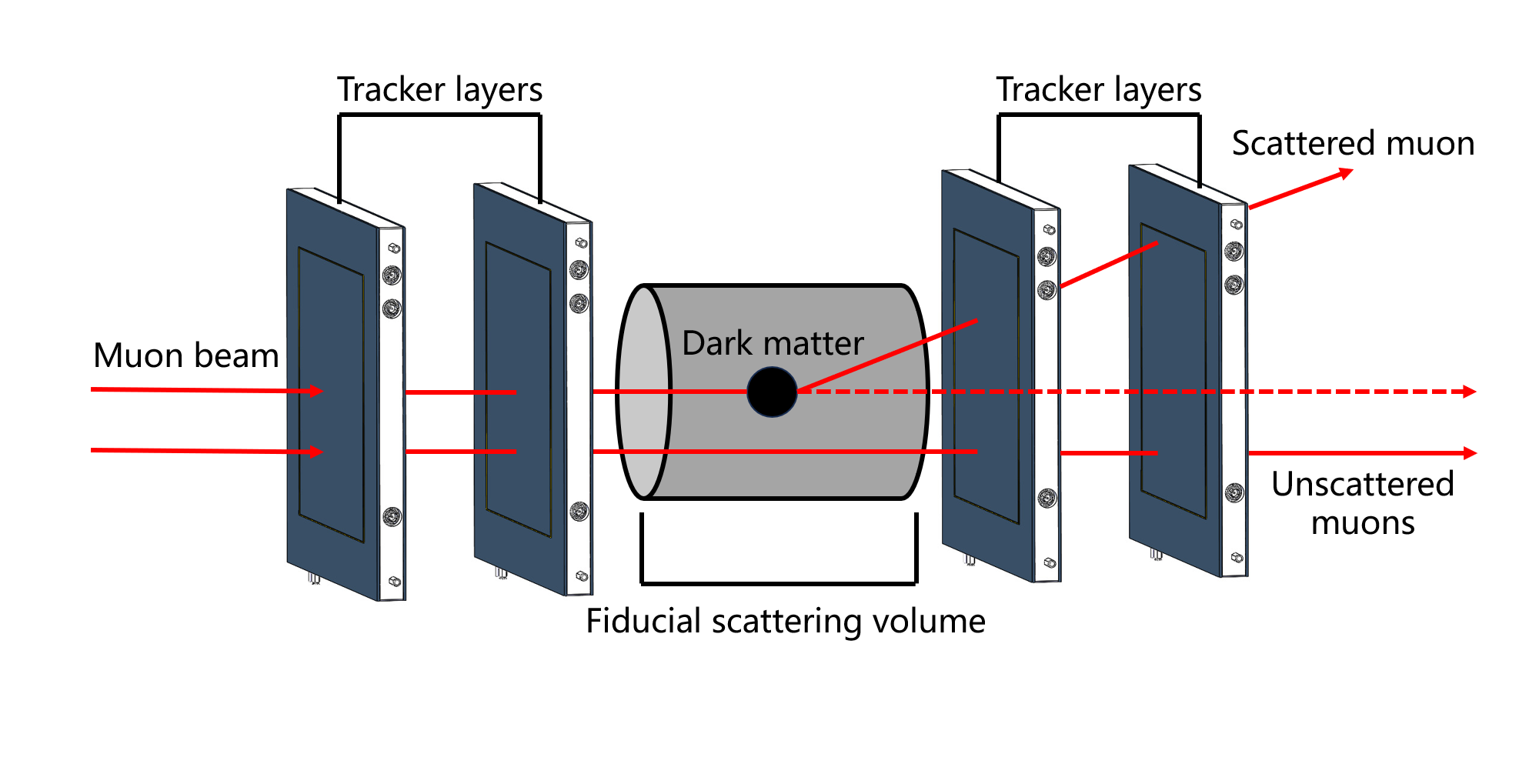}
    \captionsetup{justification=raggedright} % 局部左对齐
    \caption{Schematic illustration of the beam-muon scattering approach for probing muonphilic DM. A collimated muon beam traverses the upstream tracker layers and enters the fiducial scattering volume. If an elastic scattering between a muon and a DM particle occurs, the outgoing muon is deflected from the nominal beam direction and reconstructed by the downstream tracker layers. Unscattered muons continue along the beam axis and define the dominant background reference distribution.}
    \label{fig:1}
\end{figure}

However, the cosmic-ray method is intrinsically limited by the low natural muon flux, the broad energy and angular distributions, and the presence of nonmuon secondary particles. A dedicated accelerator muon beam can overcome these limitations by providing a much higher intensity, a controlled incident direction, and a well-defined beam phase space. As illustrated schematically in Fig.~\ref{fig:1}, the beam-muon approach replaces the stochastic cosmic-ray source with a collimated muon beam traversing a fiducial scattering volume. The upstream and downstream tracking detectors reconstruct the incoming and outgoing trajectories, and candidate $\mu$-DM scattering events are searched for through deviations in the angular distribution. In this configuration, the sensitivity is expected to scale primarily with the total number of incident muons and the effective path length inside the fiducial volume, rather than being strongly determined by the transverse beam spot size.

In this work, we present a simulation study of the prospects for probing slow muonphilic DM with accelerator muon beams. A Geant4-based framework \cite{AGOSTINELLI2003250} is developed to model muon transport, detector response, track reconstruction, and elastic $\mu$-DM scattering signal. We first investigate ideal pure-muon beams to quantify the dependence of the projected sensitivity on beam energy, transverse beam size, and exposure time. For a benchmark beam intensity of $10^5/\rm{s}$, the projected sensitivity surpasses the previous cosmic-ray result within approximately 11 s and improves it by about two orders of magnitude after a one-day exposure. We further implement a realistic muon-beam phase-space distribution based on simulations for the HIAF \cite{PhysRevAccelBeams.28.053401, An_2025} and find that the expected upper limits are comparable to those obtained with an ideal pure-muon beam. These results indicate that a beam-muon scattering experiment can provide a robust and experimentally feasible route toward significantly improved sensitivity to slow muonphilic DM.

\section{Principle}

Following the experiment in Ref.~\cite{5jh7-fxf4}, the expected signal rate for muon-DM elastic scattering in a beam-based configuration can be expressed as

\[
N_{sig} = \frac{\rho_{\rm DM} \times f_{E}}{m_{\rm DM}} \, \sigma_{\mu,\rm DM} \, F_\mu \, t \, \epsilon \, \Omega_a \, V
\tag{1}
\]

where $V$ is the sensitive volume. For a beam-based configuration, the volume can be factored as $V = A \cdot L$, where $A$ is the transverse area and $L$ is the effective path length through the detection system. The muon flux $F_\mu$ (per unit area per unit time) multiplied by the area $A$ gives the total beam intensity $I_\mu = F_\mu A$. The signal rate thus becomes

\[
N_{sig} = \frac{\rho_{\rm DM} \times f_{E}}{m_{\rm DM}} \, \sigma_{\mu,\rm DM} \, I_\mu \, t \, \epsilon \, \Omega_a \, L
\tag{2}
\]

In this form, the transverse beam size does not appear explicitly; it is absorbed into the beam intensity term. As long as the beam spot is fully contained within the detector acceptance, the sensitivity is determined by the total muon intensity and the path length through the fiducial volume, rather than the specific transverse profile. Consequently, the resulting cross-section limits are expected to be largely insensitive to the beam spot size. The dependence on beam energy, on the other hand, enters through scattering kinematics and reconstruction efficiency, and its impact requires quantitative evaluation through simulations.

\section{Simulation}

The simulation framework is developed based on the PKMu Geant4 codebase~\cite{PKMUON_2024}, which provides the detector geometry, material configuration, and physics lists for the muon tomography system. The detector geometry and all standard physics processes are identical to those described in Ref.~\cite{Zhang2026}. The detector contains four RPC tracking layers.
Each layer has a 28 × 28 cm$^2$ active area. The nominal
interlayer spacings are 20, 50, and 20 cm.

A dedicated class is implemented to model elastic scattering between muons and DM particles. The DM is treated as a non-relativistic component with a Maxwellian velocity distribution characterized by $v_{\rm DM}\sim220\,\mathrm{km/s}$ and local density $\rho_{\rm DM}=0.3\,\mathrm{GeV/cm^3}$~\cite{Yu2024}. The interaction is parameterized phenomenologically with the cross section $\sigma_{\mu,\mathrm{DM}}$ as a free parameter. The scattering probability along each muon step is computed as $P = (\rho_{\rm DM}/m_{\rm DM})\,\sigma_{\mu,\mathrm{DM}}\,\Delta l$, implemented as a discrete stepping probability within Geant4.

\begin{figure*}[htbp]
  \centering
  \begin{subfigure}[b]{0.48\textwidth}
    \includegraphics[width=\linewidth]{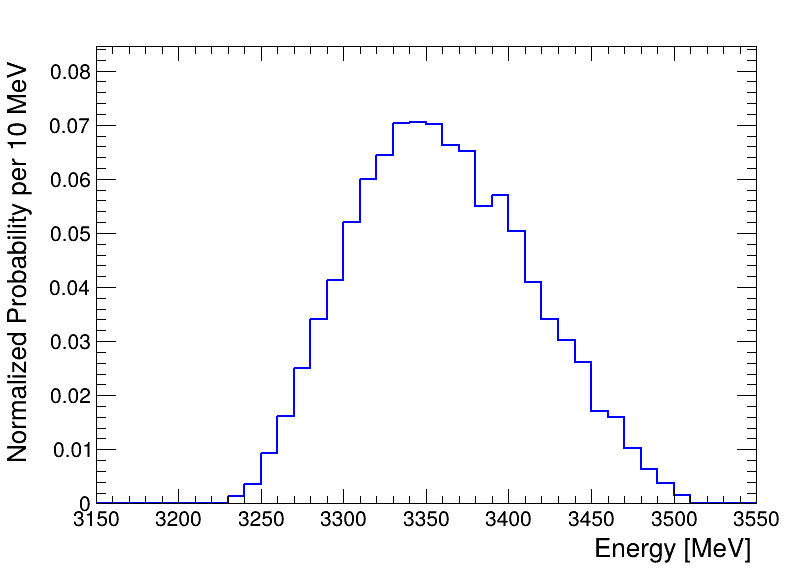}
  \end{subfigure}
  \hfill
  \begin{subfigure}[b]{0.48\textwidth}
    \includegraphics[width=\linewidth]{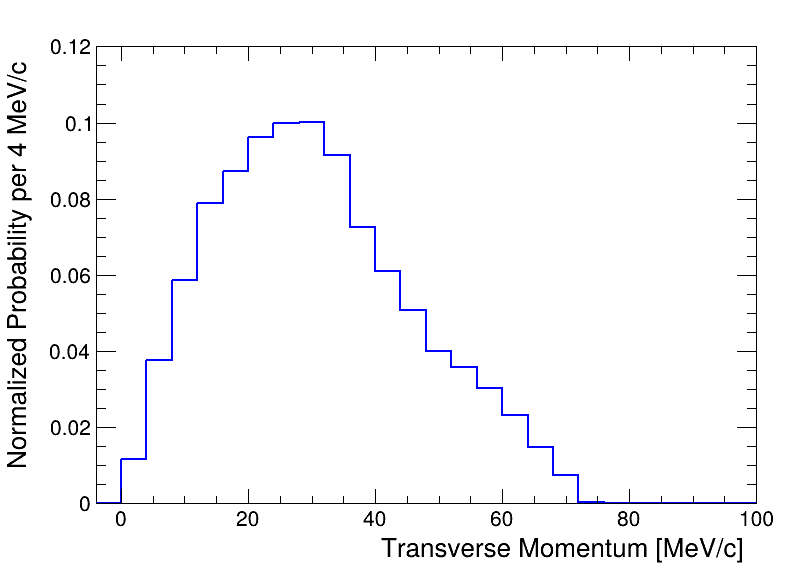}
  \end{subfigure}
  \caption{(a) Energy distribution and (b) transverse momentum distribution of muons from the HIAF beam-line simulation.}
  \label{fig:spectra}
\end{figure*}

\begin{figure}[htbp]
    \centering
    \includegraphics[width=1.0\columnwidth]{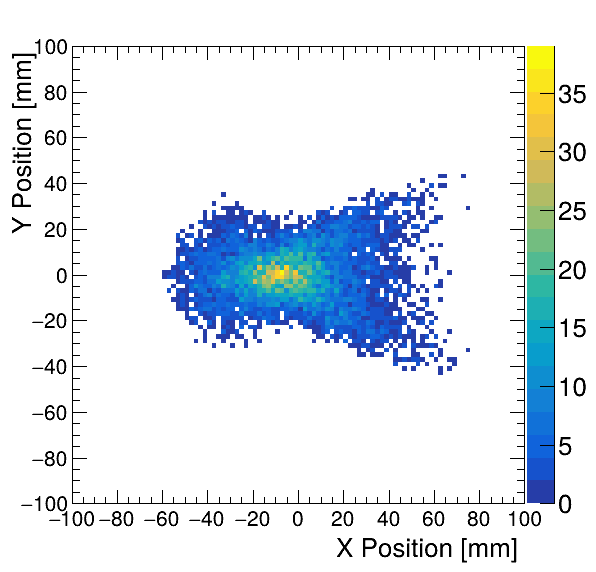}
    \caption{Transverse position distribution of muons at the entrance plane from the HIAF beam-line simulation.}
    \label{fig:xy}
\end{figure}

The particle source is positioned on the top surface plane of the detection system. Two muon source configurations are implemented. For the first configuration, muons are generated with fixed energy and momentum, vertically incident on the detection system. For the second configuration, the particle type, initial position, and four-momentum are sampled from the HIAF simulation output~\cite{PhysRevAccelBeams.28.053401}. A total of approximately $8\times10^3$ muons are generated, from which the beam phase-space parameters are estimated and the distributions in Figs.~\ref{fig:spectra} and~\ref{fig:xy} are obtained. The energy of particles is concentrated around 3.3-3.5 GeV, while the transverse momentum is mainly below about 80 MeV/c, and the purity of muons is approximately 99.8\%.

To evaluate the angular acceptance, $1.1\times10^6$ muons are generated for each Gaussian source configuration under two triggering conditions, namely requiring coincidences from all four detector layers and requiring coincidences from only the top two layers. The same method is used in the HIAF source for different DM mass configurations.

\section{Results and Discussion}

For each configuration described above, the expected upper limit on the muon-DM scattering cross section is derived following the statistical procedure established in Ref.~\cite{5jh7-fxf4}, using the CL$_s$ method implemented in the Higgs combine tool~\cite{JUNK1999435, Read_2002}. Unless otherwise specified, all results in this section are presented for $m_{\rm DM}=1$ GeV as a representative benchmark.

The dependence on the transverse beam size is first examined. Pure muon beams with fixed energy ($E_\mu=1$ GeV) are generated with Gaussian transverse profiles of varying widths: $\sigma_{beam} = 1.5$, 2, and 3 cm. The resulting $95\%$ CL upper limits on the scattering cross section are shown in Table~\ref{tab:spotsize}. The limits remain nearly constant across all configurations, with variations well within the statistical uncertainties, confirming that the limits are insensitive to the beam spot size when the beam is fully contained within the detector acceptance. To further investigate the dominant factor affecting the sensitivity, the detector geometry is modified by increasing the interlayer spacings from 20-50-20 cm to 20-100-20 cm. As shown in the same table, this change leads to an improvement in the upper limit, indicating that the detector geometry is a more significant factor than the beam spot size.

\begin{table}[htbp]
\centering
\caption{95\% CL upper limits on $\sigma_{\mu,\rm DM}$ for different beam spot sizes and detector spacings ($E_\mu=1$ GeV, $m_{\rm DM}=1$ GeV, $N_{\rm events} = 1.1\times10^6$).}
\begin{tabular}{c c c}
\hline
$\sigma_{beam}$ (cm) & Spacing (cm) & $\sigma_{\mu,\rm DM}$ (cm$^2$) \\
\hline
1.5 & 20-50-20 & $(2.7\pm1.1)\times 10^{-18}$ \\
2 & 20-50-20 & $(2.7\pm1.3)\times 10^{-18}$ \\
3 & 20–50–20 & $(2.9\pm1.3)\times 10^{-18}$ \\
1.5 & 20–100–20 & $(1.0\pm0.5)\times 10^{-18}$ \\
\hline
\end{tabular}
\label{tab:spotsize}
\end{table}

\begin{figure}[htbp]
    \centering
    \includegraphics[width=0.85\columnwidth]{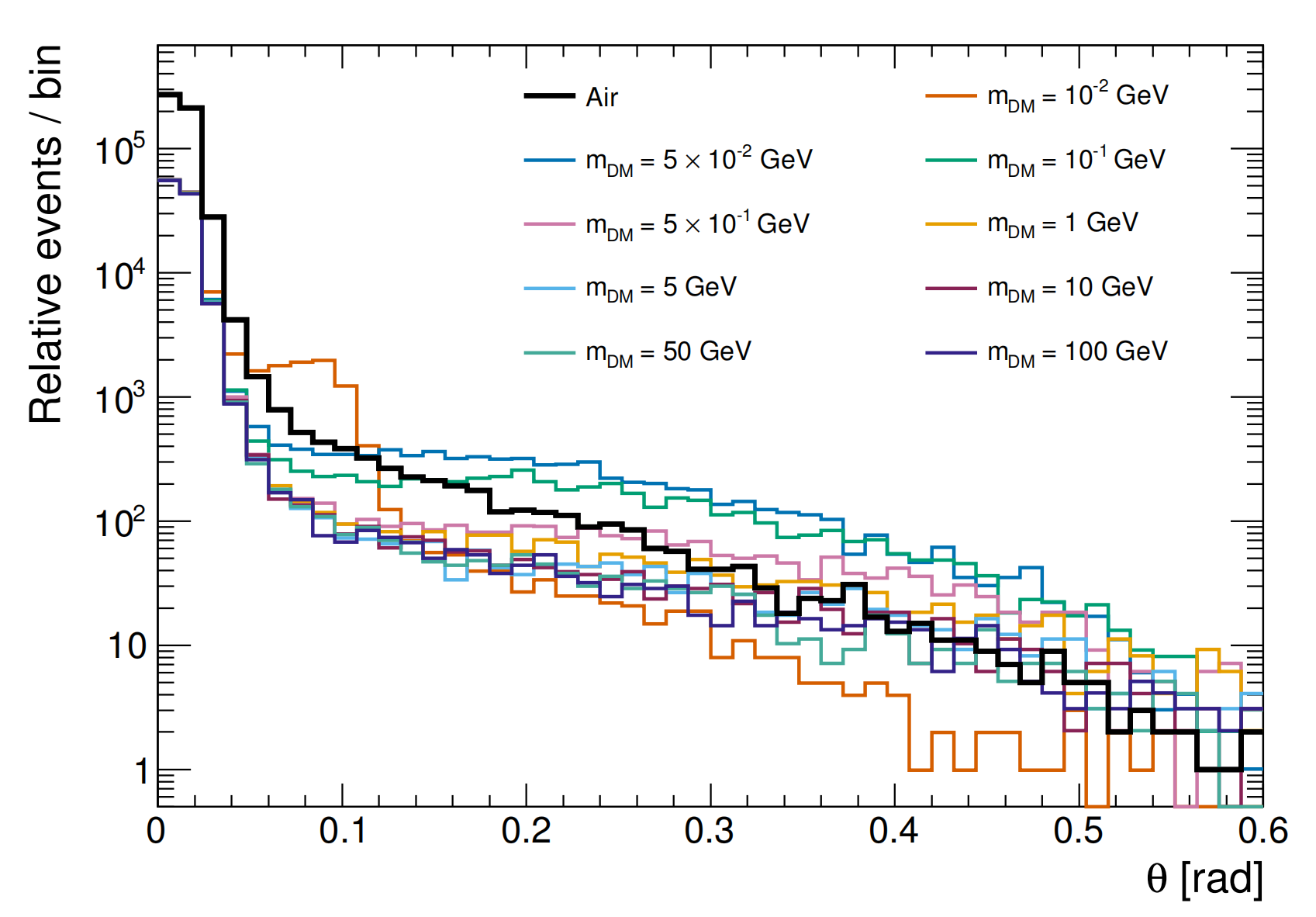}
    \caption{Normalized scattering angle distributions for the pure muon beam with $E_\mu = 1$ GeV and $\sigma_{beam}=3$ cm transverse profile. The background-only (Air) distribution is fixed with $N$ events, and signal distributions are scaled by $N/n$, where $n$ is the number of events in each signal sample. The selection requiring the incident point to be within a transverse radius of 3 cm is applied. The figure is shown for illustrative purposes only.}
    \label{fig:theta_example}
\end{figure}

As an illustration of the typical scattering angle distributions obtained in the simulation, Fig.~\ref{fig:theta_example} shows the normalized angular distributions for the pure muon beam with $E_\mu = 1$ GeV and a Gaussian transverse profile of $\sigma_{beam} = 3$ cm, along with the background-only (Air) distribution. The air distribution is kept at its original normalization with $N$ events, while each signal distribution is scaled by a factor of $N/n$, where $n$ is the number of events in the corresponding signal sample. A selection requiring the incident point to lie within a circle of radius 3 cm in the transverse plane is applied.

\begin{table}[htbp]
\centering
\caption{95\% CL upper limits on $\sigma_{\mu,\rm DM}$ for different beam energies ($m_{\rm DM}=1$ GeV, Gaussian profile with $\sigma_{beam}=3$ cm, $N_{\rm events} = 1.1\times10^6$).}
\begin{tabular}{c c}
\hline
$E_\mu$ (GeV) & $\sigma_{\mu,\rm DM}$ (cm$^2$) \\
\hline
1 & $(2.9\pm1.3)\times 10^{-18}$ \\
3 & $(3.3\pm1.4)\times 10^{-18}$ \\
5 & $(3.4\pm1.4)\times 10^{-18}$ \\
\hline
\end{tabular}
\label{tab:energy}
\end{table}

The dependence on beam energy is investigated next. Pure muon beams with fixed energy of 1, 3, and 5 GeV are generated with the identical transverse profile ($\sigma_{beam}=3$ cm) and the standard 20-50-20 cm spacing. The limits are also found to remain constant, indicating that the sensitivity is only weakly dependent on the incident energy for relativistic muons in the GeV range. The detailed results are listed in Table~\ref{tab:energy}.

\begin{table*}[htbp]
\centering
\caption{95\% CL upper limits on $\sigma_{\mu,\rm DM}$ for different DM masses using the HIAF muon source. The calculated limits correspond to $N_{\rm events}\approx8\times10^3$, while the expected limits correspond to $N_{\rm events}=8.64\times10^9$ (one day exposure with a beam intensity of $10^5$ s$^{-1}$).}
\begin{tabular}{c c c}
\hline
$m_{\rm DM}$ (GeV) & $\sigma_{\mu,\rm DM}$ (cm$^2$, calculated) & $\sigma_{\mu,\rm DM}$ (cm$^2$, expected) \\
\hline
$1.0\times 10^{-2}$ & $(3.1\pm1.4)\times 10^{-19}$ & $(8.6\pm3.8)\times 10^{-22}$\\
$5.0\times 10^{-2}$ & $(9.2\pm6.4)\times 10^{-19}$ & $(2.5\pm1.8)\times 10^{-21}$\\
$1.0\times 10^{-1}$ & $(2.0\pm1.3)\times 10^{-18}$ & $(5.6\pm3.6)\times 10^{-21}$\\
$5.0\times 10^{-1}$ & $(1.3\pm0.8)\times 10^{-17}$ & $(3.7\pm2.2)\times 10^{-20}$\\
$1.0\times 10^{0}$ & $(2.8\pm1.6)\times 10^{-17}$ & $(7.7\pm4.4)\times 10^{-20}$\\
$5.0\times 10^{0}$ & $(1.6\pm0.8)\times 10^{-16}$ & $(4.5\pm2.3)\times 10^{-19}$\\
$1.0\times 10^{1}$ & $(3.2\pm1.6)\times 10^{-16}$ & $(8.8\pm4.6)\times 10^{-19}$\\
$5.0\times 10^{1}$ & $(1.7\pm0.9)\times 10^{-15}$ & $(4.7\pm2.4)\times 10^{-18}$\\
$1.0\times 10^{2}$ & $(3.6\pm1.8)\times 10^{-15}$ & $(9.9\pm4.6)\times 10^{-18}$\\
\hline
\end{tabular}
\label{tab:hiaf}
\end{table*}

\begin{figure}[htbp]
    \centering
    \includegraphics[width=1.0\columnwidth]{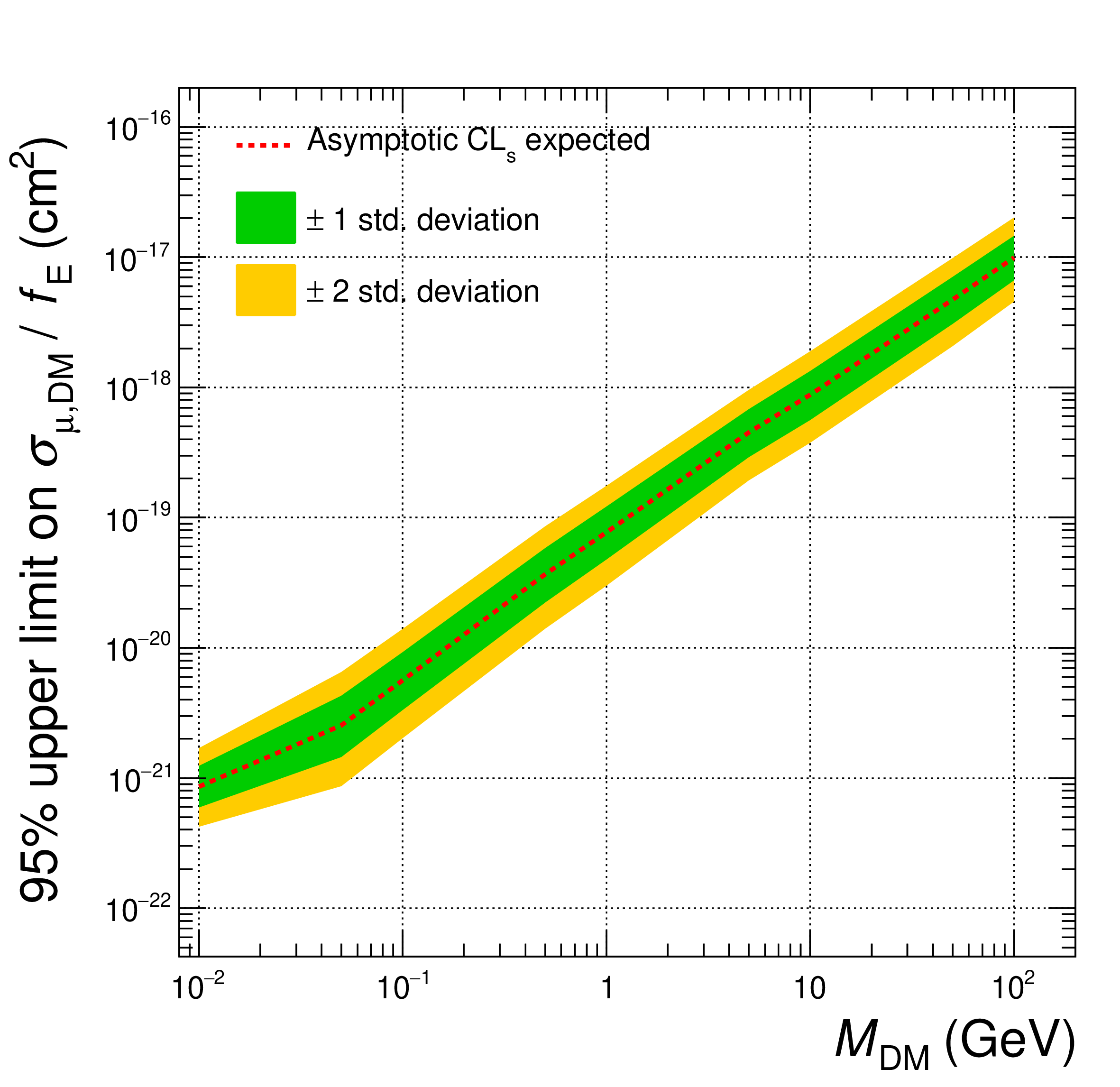}
    \caption{Expected $95\%$ CL upper limits on $\sigma_{\mu,\rm DM}$ as a function of $m_{\rm DM}$ for the HIAF muon source with a one-day exposure ($I_\mu=10^5$ s$^{-1}$).}
    \label{fig:hiaf_limit}
\end{figure}

In HIAF source simulation, the particle information---including particle type, initial transverse position, and four-momentum---is then sampled by the Geant4 particle gun and propagated through the simulation framework described above. For each DM mass hypothesis, the resulting $95\%$ CL upper limits on $\sigma_{\mu,\rm DM}$ are listed in the second column of Table~\ref{tab:hiaf}. The third column lists the expected limits for a one-day run with a beam intensity of $10^5$ s$^{-1}$. These limits are presented as a function of $m_{\rm DM}$ in Fig.~\ref{fig:hiaf_limit}. As expected, the beam-based limits comfortably surpass the cosmic-ray limit from the 63-day measurement in Ref.~\cite{5jh7-fxf4} by approximately two orders of magnitude, while maintaining the same trend in mass dependence.

\section{Summary and Outlook}

The simulation results demonstrate that the projected sensitivity is largely insensitive to both the transverse beam size and the incident muon energy, provided the beam spot is fully contained within the detector acceptance. In contrast, the detector geometry---specifically the spacing between tracking layers---is identified as the dominant factor affecting the sensitivity. For a benchmark beam intensity of $10^5$ s$^{-1}$, the simulated pure-muon beam surpasses the cosmic-ray limit of $1.61\times10^{-17}$ cm$^2$ at $m_{\rm DM}=1$ GeV within approximately 11 seconds. The HIAF muon source, simulated with a realistic beam phase-space distribution, yields projected limits that improve upon the existing cosmic-ray results by nearly two orders of magnitude in a one-day exposure.

This work further benefits from the recent developments at HIAF in Huizhou, China. The muon production has recently been achieved at the facility\footnote{The beam test results are under study and will be reported elsewhere.}. Time-of-flight particle identification has been performed, and muon imaging of test samples has been carried out, demonstrating the feasibility of muon-related applications at HIAF. A joint muon spectrometer has been constructed at the beam line by the Institute of Modern Physics (IMP) in collaboration with the PKMu group. The spectrometer integrates resistive plate chambers, drift chambers, and other detector subsystems, providing a versatile platform for muon beam characterization and a broad physics program, including the beam-based $\mu$-DM scattering experiment proposed in this work.

Looking forward, extended running with increased beam intensity and optimized detector configurations could push the sensitivity further, potentially probing unexplored regions of the muonphilic DM parameter space and complementing the ongoing efforts at NA64$\mu$ and other accelerator-based dark sector searches. 

\section{Acknowledgments}

This work is supported in part by the National Natural Science Foundation of China under Grant No.~12325504.

% The \nocite command causes all entries in a bibliography to be printed out
% whether or not they are actually referenced in the text. This is appropriate
% for the sample file to show the different styles of references, but authors
% most likely will not want to use it.
\nocite{*}

\bibliography{apssamp}% Produces the bibliography via BibTeX.
\bibliographystyle{apsrev4-2}

\end{document}